\documentclass{mn2e}
\usepackage{epsfig}
\usepackage{pslatex}

\def\msun{{\rm M_{\odot}}}

\def\today{\number\year \ \ifcase\month\or
  January\or February\or March\or April\or May\or June\or
  July\or August\or September\or October\or November\or December
 \fi \ \number\day }
\date{Accepted ??. Received ??; in original form \today}

\volume{000}

\pagerange{\pageref{firstpage}--\pageref{lastpage}} \pubyear{2004}

\begin{document}

\label{firstpage}

\title[Wolf-Rayet and O Star Runaway Populations from Supernovae]
{Wolf-Rayet and O Star Runaway Populations from Supernovae}

\author[L.\,M. ~Dray, J.\,E. ~Dale, M.\,E.~Beer, R.~Napiwotzki,
  A.\,R.~King]{L.\,M. ~Dray\thanks{E-mail: Lynnette.Dray@astro.le.ac.uk}, J.\,E. ~Dale, M.\,E.~Beer, R. ~Napiwotzki, 
A.\,R.~King\\ 
Theoretical Astrophysics Group, University of Leicester,
Leicester, LE1~7RH, UK\\
}

\maketitle

\begin{abstract}
We present numerical simulations of the runaway fractions expected amongst 
O and Wolf-Rayet star populations resulting from stars ejected from binaries by the 
supernova of the companion. Observationally the runaway fraction for both types of 
star is similar, prompting the explanation that close dynamical interactions are 
the main cause of these high-velocity stars. We show that, provided that the initial 
binary fraction is high, a scenario in which two-thirds of massive runaways are from 
supernovae is consistent with these observations. Our models also predict a low 
frequency of runaways with neutron star companions and a very low fraction of 
observable Wolf-Rayet--compact companion systems.

\end{abstract}

\begin{keywords}
binaries: close -- stars: early-type -- stars: kinematics -- stars: Wolf-Rayet -- supernovae: general
\end{keywords}

\section{Introduction}
OB runaway stars are massive, early-type stars which have high peculiar 
velocities relative to the local standard of rest. These may reach values 
as large as $200 \, {\rm kms}^{-1}$, though most are below $100 \, {\rm kms}^{-1}$. The 
lower velocity limit for a star to be considered a runaway varies 
between studies but is generally between $20$ -- $40 \,{\rm kms}^{-1}$. 
There are also a number of early-type stars at high Galactic latitudes 
which require high velocities if they were formed in the plane 
in order to reach their current locations within their lifetimes (Conlon et al. 1992, Allen \& Kinman 2004).
In this paper we take the threshold velocity for a 
star to be deemed runaway to be $30 \, {\rm kms}^{-1}$. 
Despite their high masses, 10 -- 30 \%
of O stars and 5 -- 10 \% of B stars have runaway status (Gies 1987).   

There are two
likely ways in which such relatively massive stars can acquire high
velocities. First, they may have been members of close binary systems
which were disrupted by the supernova explosion of the companion 
(the Binary Supernova Scenario (BSS), Blaauw 1961). In
this case the runaway velocity must be similar to the star's orbital
velocity before the supernova. Second, runaway velocities can arise
because the star interacted dynamically with members of its natal star
cluster (the Dynamical Ejection Scenario (DES), Poveda, Ruiz \& Allen 1967). As early--type 
runaway stars are fairly massive, this
suggests that the interaction was with a binary system of two massive
stars, and the most likely process is binary--binary scattering, in
which the eventual runaway was a member of a binary which was
disrupted by a more massive one. Clear examples of both types of of
runaway (supernova disruption and dynamical interaction) are known
(Hoogerwerf, de Bruijne \& de Zeeuw 2000). The fraction of
runaways originating from either 
route is less clear; significant numbers of runaways do show
signs of interaction with a close companion, suggesting that
supernovae must be implicated for a substantial fraction. However, 
high-latitude early-type stars have normal rotational velocities for 
their stellar type (Lynn et al. 2004) as opposed to the fast rotation which might be 
expected for the secondary from an interacting binary.

Hoogerwerf, de Bruijne \& de Zeeuw (2001) trace the paths of runaway stars 
back to their parent clusters, and find perhaps two thirds are 
the result of supernova ejection. Blaauw (1993) 
finds that over 50\% of massive runaways have enhanced surface He abundances
and high rotational velocities, 
suggesting that their parent systems experienced mass transfer and therefore 
are good candidates for supernova separation. It should be noted that rapid 
rotation and enhanced abundances are not in themselves 
unambiguous signs of accretion having taken place, since if a star is formed 
with rapid rotation then this will affect the surface abundances over its 
lifetime (e.g. Fliegner, Langer \& Venn 1996); however the high proportion 
of runaways displaying these properties suggests that they are related in these cases to the 
circumstances which make a runaway. This favours the BSS, which is expected 
to make such stars. Whilst 
theoretically dynamical ejection could occur in an interacting system after it has 
undergone mass transfer, producing a DES runaway with BSS-expected abundances, it it is more likely to have occurred on the
main sequence before mass transfer; furthermore, as tighter systems, those with early interaction present 
a smaller cross-section to collision and are harder to unbind.

On the other hand, population synthesis suggests that only a relatively small fraction of 
O star runaways can have come about via the BSS (e.g. Portegies Zwart 2000).

The two pictures sketched above predict in principle what kinds of
stars become runaways. 
However the O and/or B phases are only the early stages of the life of a massive 
star; in particular, many massive stars at solar metallicity go on to become 
Wolf-Rayet (WR) stars (Chiosi \& Maeder 1986). WR stars are characterised by small or absent H 
envelopes as a result of high mass loss; therefore they generally arise from the most 
massive stars, which have higher mass-loss rates throughout their lives. 
In a binary this process may be affected by Roche Lobe Overflow (RLOF), both as the donor (where the mass loss 
strips the envelope) and the gainer (which may become massive enough, via accretion, 
to later undergo a WR phase).

Some proportion of WR stars would therefore also 
be expected to be runaway, with the exact numbers depending on the dominant method 
of runaway production.
Recently evidence has emerged that the fraction
of runaways among Wolf-Rayet stars is similar to that amongst O stars (Mason et al. 1998, Moffat et al. 1998,
Foellmi et al. 2003) at around 10 per cent. This has been interpreted as evidence in favour of dynamical 
ejection being the primary route for making massive runaway stars.
In this paper we consider whether the SN route may also produce similar runaway fractions
for O and WR stars, and examine the implications of this.

\section{Runaways via the Dynamical ejection scenario}
The dynamical ejection scenario was first proposed by Poveda et al. 
(1967). It involves binary encounters in the cores of the most
massive OB associations. These encounters extract energy from the
binary orbit through tightening of the orbit and generate runaway
stars.

One system is known in which there is strong evidence for an origin
through the dynamical ejection scenario. AE\,Aur and $\mu$\,Col are
both O9.5V stars with similar ages which 
are running away in opposite directions with velocities relative to
the local standard of rest of 113.3 and 107.8\,kms$^{-1}$ respectively
(Hoogerwerf et al. 2001). Because of their similar
but oppositely directed space velocities Blaauw \& Morgan (1954) first
proposed a common origin for these runaway stars in the Orion
nebula. Gies \& Bolton (1986) proposed a binary-binary interaction
formed both of these runaway stars and the binary
$\iota$\,Ori. Supporting this hypothesis Hoogerwerf et al. (2001) have
shown that these three objects occupied a very small region of space
2.5\,Myr ago in the Trapezium cluster. Gualandris, Portegies Zwart \& Eggleton
(2004) have performed N-body simulations of the binary-binary
encounter and have shown how a binary-binary encounter with an exchange
occurring between the two binaries could produce the currently observed
configuration.

Hoogerwerf et al. (2001) investigated the origin of twenty-two nearby
runaway stars. Parent associations were proposed for sixteen of
these stars. Of these sixteen eleven were proposed to have been
produced via the binary supernova scenario as opposed to five through
the dynamical ejection scenario. Two of the remaining runaways had
more than one possible parent associations but were consistent with an
origin via the binary supernova scenario. This implies a fraction of 
runaways formed through the dynamical ejection scenario as less than 
one third. Leonard (1991) has performed a number of binary-binary
encounters and finds that runaway velocities are greatest when the
initial binaries are circular and the stars have similar masses. For
the most-massive runaways in these binary-binary encounters Leonard
(1991) finds that the maximum runaway velocity is half the surface
escape velocity of that star - so escape velocities of 100s of
kms$^{-1}$ are possible. 

There are a number of competing effects which will determine the relative 
sizes of the O and WR star runaway fraction in the case that all runaways are 
created by dynamical interaction. First, in a binary-single star interaction, 
the star which is most likely to be ejected as a runaway is the least massive 
of the three. It is thought that runaway O stars arise from 
binary-binary (Clarke \& Pringle 1992) or higher-order multiple  
interactions, but even here it is the less massive stars which 
will be ejected with the greatest velocities. Selecting for lower-mass 
stars selects for O stars (the initial mass limit above which a star 
goes through an O phase being lower than the initial mass limit above 
which a star goes through a WR phase) and for evolved WR stars 
(if the interaction happens late in the lifetime of the WR star when 
it has lost much of its mass via a wind). Once a star is a runaway it 
is likely to remain that way, which selects for the later stages in a 
star's lifetime, i.e. against O stars.

Between these considerations it 
is difficult to form a clear picture of the population occurring 
from dynamical interactions without large-scale numerical simulations. 
If the runaway fractions of such WR and O stars are equal, it may be 
only because of the balance of competing effects.

\section{Runaways via Supernovae}

Even if it is completely symmetric, a supernova in a binary system still occurs 
away from the centre of mass of the system and hence imparts a net velocity 
to the system as a whole. Whether the system is unbound by this and whether 
the resulting velocity of the companion is great enough for it to be observed 
as a runaway depends on its pre-SN parameters. In particular, a binary 
will remain bound despite a SN explosion if less than half its total mass is lost. For most  
binaries which have undergone mass transfer, it is the primary -- initially 
the most massive star -- which explodes first, but by the time of its SN its mass 
is less than that of its companion. Therefore for perfectly symmetric SNe all of 
these systems would remain bound. In these circumstances it is hard to explain 
the very low binary fraction amongst runaway O stars (Mason et al. 1998).

Studies of single pulsar velocities (Lyne \& Lorimer 1994) find that an additional `kick' 
velocity of some $450\,{\rm kms}^{-1}$ (imparted by asymmetric mass 
loss or neutrino emission) is required to account for the extremely high 
velocity of some neutron stars. Brandt \& Podsiadlowski (1995) suggest 
that kicks of this magnitude will unbind the binary in over 70 \% of cases.
In this case it is quite easy to make massive runaways via the BSS.
However, whilst the populations which go on to form runaway O and runaway WR 
stars are similar, they are not identical and in particular they are affected 
differently by the time at which the SN occurs and the size of the kick.

The simplest scenario in which O and WR runaway stars both arise from 
supernovae is when all kicks are equal. The properties of the runaway 
are determined by any binary interaction it may have undergone 
and its state of evolution at the time of the SN of its companion.
The fraction of WR stars which are runaways should in this case
always be significantly greater than the fraction of O stars which 
are runaways.

This arises from the relative positions of the O and WR phases in the 
star's lifetime. Stars either begin their lives as O stars or become O 
stars after undergoing accretion from RLOF. However, the O phase is always 
earlier than the WR phase. In a binary it is reasonable to assume that 
both stars are formed at the same time. By the time one of the stars has 
evolved to the point that it undergoes a supernova, the other (which 
will become the runaway) has already gone through some or all of its 
O star phase, but is unlikely to have become a WR star yet. The statistics 
of observed WR binaries back this up: of 20 Galactic WR binaries with 
measured masses (van der Hucht 2001), at least 14 have as companions 
O stars which are massive enough to subsequently go through a WR phase, 
but only one (WR20a, which, as a system of an $83 \, \msun$ and an 
$82 \, \msun$ star (Bonanos et al. 2004), is highly unusual) 
contains two WR stars. 

One potential way of getting around this is to invoke a population with 
very uneven mass ratios, such that the less massive star still 
has most of its O star lifetime left after the SN of its 
companion. However, in order to be an O star on the main sequence at solar metallicity 
a mass of at least $17 \msun$ is required. This then requires 
the other star to be extremely massive, and so is unlikely to be a common situation.
Furthermore, the masses of the components in massive binaries are correlated (Garmany, Conti \& 
Massey 1980), with very few unevolved systems having mass ratios $q = M_{2}/M_{1}$ below 0.3.

However, the above assumes that all supernovae are equal -- that is, that 
the distribution of parameters which produce WR evolution in the 
secondary\footnote{NB. In this paper we will follow the theorists' convention of 
referring to the {\it initially} most massive star as the primary throughout, 
even if through interaction or winds it becomes the less massive star in the system.}
and the distribution of parameters which produce an 
asymmetric SN explosion which unbinds the system are completely unrelated. 
In reality this is unlikely to be the case.

The likelihood of a system producing a runaway WR star depends on a number 
of factors. First, the star which survives beyond the SN of its companion
must be, or become, massive enough to undergo a WR phase. In the case 
that mass transfer occurs, the mass limit for a star to go through a WR 
phase is lowered somewhat (Dray \& Tout 2005) due to accretion of 
He-enhanced matter and subsequent 
thermohaline mixing. Mass transfer also promotes the likelihood of the 
secondary undergoing a WR phase by raising its mass, although 
the amount of mass transfer which can occur before the secondary is spun 
up to rotational break-up velocity is a matter of debate (Packet 1981, Dewi 2005).

Second, this star must then be given a large enough velocity by the SN of 
the other that it is observable as a runaway. Here we use $30 {\rm kms}^{-1}$
as the velocity threshold above which a star is classified as runaway. Recent work by Pfahl et al. 
(2001) suggests that the velocity distribution of observed X-ray binary systems is 
indicative of a bimodal distribution of SN kicks, with smaller kicks originating 
from systems which underwent mass transfer earlier in their evolution. Early mass 
transfer (i.e. short initial period) has also been linked with more nearly 
conservative mass transfer (Langer 2005). In addition, simulations of WR production 
in these systems (Dray \& Tout 2005) suggest that early mass transfer increases the 
likelihood of the secondary undergoing a WR phase. 

A further notable effect is that, since masses of stars in massive binaries are 
correlated, primaries of high mass tend to have relatively massive secondaries. 
Therefore, in the population as a whole, as the primary mass increases, so does 
the likelihood of the secondary being massive enough to undergo a WR phase. 
Above a primary mass of $40 \,  \msun$ or so, nearly all solar metallicity systems 
which interact and avoid merging will have secondaries which can go through a WR phase. Therefore 
the initial primary mass distribution of systems which can make WR runaways 
is strongly skewed towards high initial primary mass. However, $40 \, \msun$ is 
also the threshold above which the remaining cores of stars after losing mass in RLOF and winds 
at solar metallicity are above $8 \, \msun$ -- i.e. massive enough to collapse to a black hole 
rather than a neutron star (Fryer 1999, van den Heuvel et al. 2000). 
The level of wind mass loss for WR stars is a matter of debate, particularly when 
rotation is considered (e.g. Maeder \& Meynet 2000). However it is 
notable that, in all five of the Galactic WC+O systems with measured masses, the WC star 
(which is a stripped core nearing the very end of its lifetime) is more massive 
than $8 \, \msun$. These systems also have massive O star companions ($23$ -- $34\, \msun$, 
van der Hucht 2001) which will quite probably still be O stars at the 
SNe of their companions and later undergo a WR phase, making them prime examples 
of potential contributors to both the (BSS scenario) O and WR runaway populations.

If a neutron star (NS) is formed in the SN explosion, the rest of the mass of the pre-explosion star 
is lost. However, when a black hole (BH) is formed the amount of matter which falls back is 
not limited by the maximum NS mass. There is a peak in the system SN mass
loss for pre-SN primary core masses somewhere around the boundary between NS and BH formation 
(Fig. 1). The most massive stars form BHs at the end of their lifetimes. These are also the stars which lose most or
all of their envelopes in winds or RLOF before they explode. This again limits the mass loss in the explosion, 
since the matter external to the core has already been lost. 
Consequently, BH-forming explosions involve smaller mass loss, relative to the remaining system mass, 
than do most NS-forming explosions. If kicks are 
basically related to anisotropic mass loss in the SN explosion, as opposed to 
anisotropic neutrino emission, then a smaller amount of mass lost relative to the system mass 
should also correspond to a smaller average kick velocity. Even if the kick velocity is unchanged, the 
amount of mass lost affects the final velocities imparted to the stars as the source of the mass loss is 
not at the centre of mass of the system (see e.g. Tauris \& Takens 1998).

\begin{figure}
\vbox to80mm{\vfil
\psfig{figure=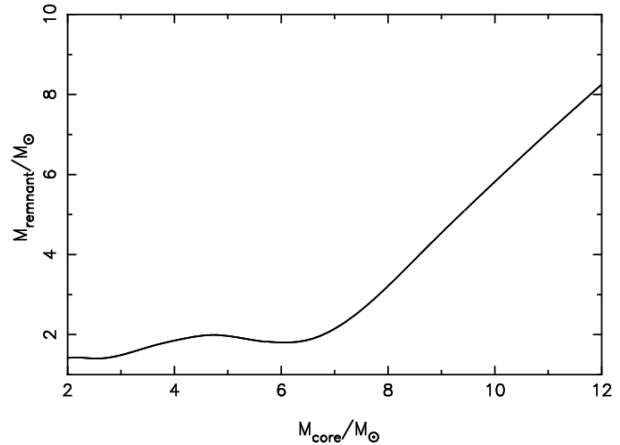,angle=270,width=80mm}
\caption{Pre-SN core mass--remnant mass relation for Z = $0.02$ from the models of 
         Woosley \& Weaver (1995) as parameterised by Portinari et al. (1998). 
         Though largely artificial in its placing of the mass cut, the true 
         relation between low-mass SNe from which all mass is lost apart from a $1.4 \,\msun$ 
         NS core up to direct-collapse BH formation from which no mass is lost is not 
         likely to be very different.}
\vfil}
\label{fig1}
\end{figure}

If the BH-forming explosion mechanism is basically similar to that forming 
a NS the imparted velocity to the companion to a BH-forming star, therefore, is still on average lower. Is this 
effect, combined with the greater likelihood of BH-forming stars to have companions which will 
become WR stars, enough to reduce the proportion of WR runaways? To test this proposition we 
use the analytic formulae provided by Tauris \& Takens (1998) for the effect of a given 
kick magnitude and direction on the companion's velocity. To quantify the effects of kicks we also 
need to know (or guess) a suitable distribution of parameters for systems immediately before 
the first SN. This in turn depends on the nature of mass transfer.

Whilst there are 
initial parameter distributions for binaries which are widely used, the masses and periods immediately 
before the SN depend critically on a number of rather less well-known evolution-based quantities, including 
the amount of matter which may be accreted during RLOF and the efficiency of common envelope mass loss. 
Fig. 2 shows Monte Carlo simulations of sets of 50000 systems with varying assumptions about input systems 
and kicks. In all cases we assume an isotropic distribution of kick directions and a Maxwellian distribution 
of kick velocities with mean $450 \,{\rm kms}^{-1}$ (Lyne \& Lorimer 1994; Lorimer, Bailes \& Harrison 1997), and that the initial binary population 
has mass ratio $q$ and period $P$ distributed according to $P(q) \propto 1$, $P(P) \propto 1/P$ with primary 
masses appropriate to a Salpeter IMF. 
Observations of massive stars suggest that the initial binary fraction is high, quite possibly close to 100 \% (Mason et al. 1998), 
though the binary fraction amongst runaways is much lower. Therefore we consider an initial population composed 
solely of binaries. Initially-single stars are highly unlikely to become runaways by either method, so the effect of 
their inclusion would be to lower the runaway fractions of both O and WR stars.

A further consideration is the appropriate metallicity. With the recent work of Apslund et al. (2005), it now seems apparent 
that the true solar metallicity is similar to that of the solar neighborhood, i.e. closer to $0.01$ that $0.02$, the usually assumed 
'solar' value when models are calculated. It is likely that runaway O stars originate from systems with a range of metallicities, 
of which $0.02$ is towards the higher end (Daflon et al. 2001). Many of the regions in which WR stars are common have metallicities which 
are close to the old value of solar metallicity (Najarro et al. 2004), due to the greater likelihood of forming such stars at 
higher Z. In order to allow comparison with previous studies, we run our 
initial calculations assuming Z $= 0.02$. However, we consider in addition the lower value of metallicity, which is also close to the metallicity 
of the LMC, and it should be borne in mind that the true Galactic value most likely results from a range between the two.

For panel {\bf a} of Fig. 2 we 
assume a simple model of binary interaction without stellar wind mass loss in which all systems with initial periods below 3000 days interact, 
systems with initial $q$ less than 0.6 or period greater than 200 days come into contact (Pols 1994) and others 
have stable mass transfer. For stable mass transfer we assume RLOF is conservative, the primary's initial mass $M_{\rm 1,i}$
and post-RLOF mass $M_{\rm 1,p}$ are related by  
\begin{equation}
 M_{\rm 1,p} = 0.058\,M_{\rm 1,i}^{1.57}
\end{equation}
and the post-RLOF period related to the initial period by 
\begin{equation}
\frac{P_{\rm p}}{P_{\rm i}} = \left( \frac{M_{\rm 1,i}M_{\rm 2,i}}{M_{\rm 1,p}M_{\rm 2,p}} \right)^{3}
\end{equation}
(van den Heuvel et al. 2000). For contact systems we assume the primary is stripped to its core as before, the secondary's 
mass remains constant and the period is determined by the energy argument of Webbink (1984),
\begin{equation}
\left( \frac{P_{\rm p}}{P_{\rm i}} \right)^{2} =  \frac{M_{\rm 1,p} + M_{\rm 2}}{M_{\rm 1,i} + M_{\rm 2}}  \left( \frac{M_{\rm 1,p}M_{2}}{M_{\rm 1,i}}\right)^{3} \left( M_{2} + \frac{2(M_{\rm 1,i} - M_{\rm 1,p})}{\eta_{\rm CE} \lambda r_{\rm L1}} \right)^{-3} ,
\end{equation}
where we take $\eta_{\rm CE}$ to be 1.0 and $\lambda$ to be 0.5 (Pfahl et al. 2002). The ratio of the Roche lobe radius of the primary 
to the separation, $r_{\rm L1}$, is taken to be
\begin{equation}
r_{\rm L1} = \frac{0.49}{0.6 + q^{2/3}\ln{(1 + q^{-1/3})}}
\end{equation}
(Eggleton 1983). Systems are considered to merge during common envelope evolution if the secondary would overflow its Roche Lobe
at the post-CE separation and masses given above.

In order to calculate the relative observed populations of WR and O stars we also need to have an idea of the time each star 
spends in these phases, both when it is runaway and when it is not. For the toy model we assume that post-RLOF, pre-SN primaries spend 
$10^{6}$ years in a WR-like phase, and, for other stars, those above $17 \, \msun$ have an O phase of $4 {\rm x} 10^{6} \,$ 
years and above $28 \, \msun$ have a WR phase of $10^{6} \,$ years. If they have accreted significantly, these mass
limits are likely lowered somewhat (Dray \& Tout 2005).
The mass lost in the SN explosion is calculated from the pre-SN core mass-remnant mass fit of Portinari, Chiosi \& Bressan (1998) to 
the SN models of Woosley \& Weaver (1995) at Z $= 0.02$.
For cores of over $15 \, \msun$ we assume direct collapse to a BH with no SN (Fryer 1999) and hence no kick.

\begin{figure*}
\vbox to140mm{\vfil
\psfig{figure=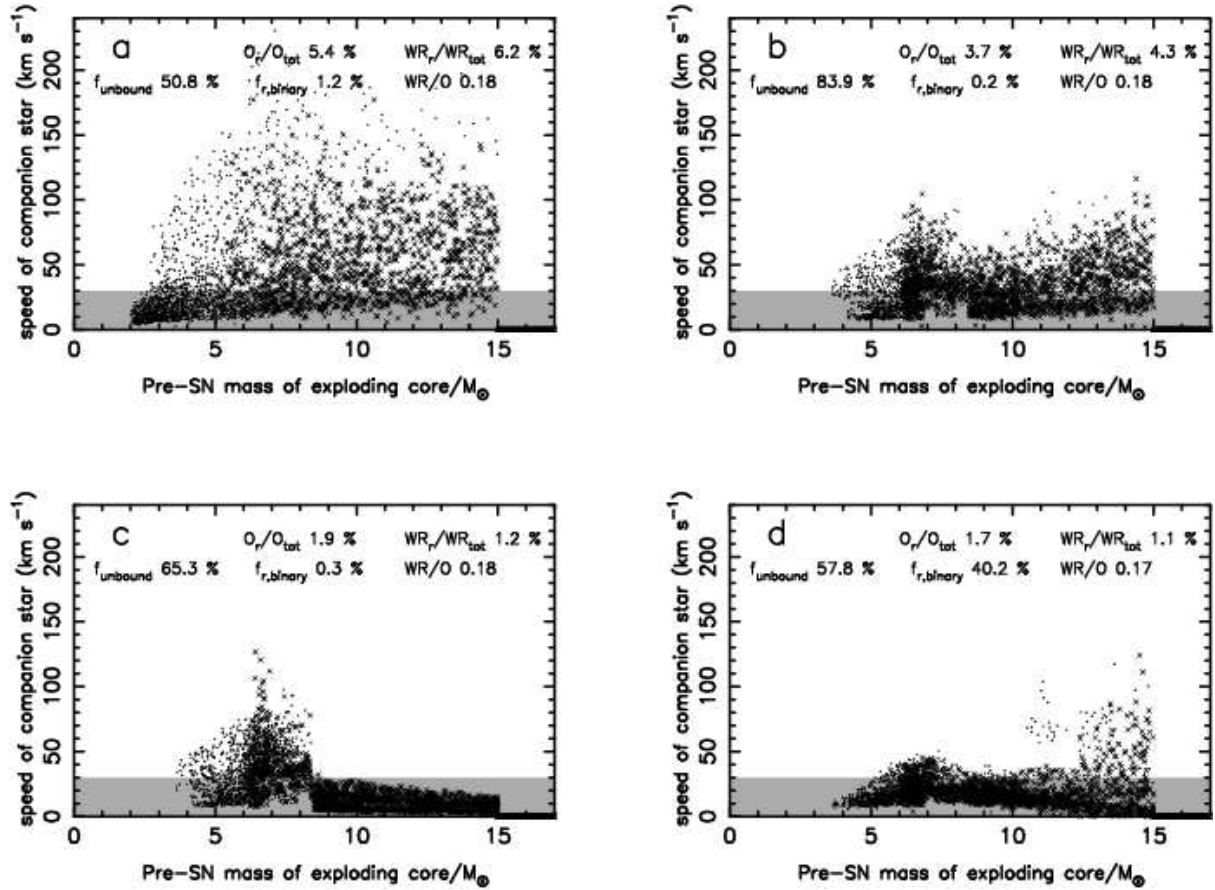,angle=270,width=160mm}
\caption{Monte-Carlo simulations of runaways resulting from massive 
         binary systems, showing the velocity of the secondary after the 
         supernova of the primary against the primary's pre-SN mass.
         The greyed-out area represents the velocity range over 
         which stars are too slow to be runaway by our definition. Solid 
         points indicate stars which go through an O star phase as runaways and crosses
         those which go through a WR phase (which may also go through an O phase).
         See text for details of the input parameters.}
\vfil}
\label{fig2}
\end{figure*}

In panels {\bf b} to {\bf d} we take pre-SN parameters from the evolutionary models of Dray \& Tout (2005)
for binaries at Z = $0.02$. We use the non-conservative RLOF set of models for which an accreting star can only 
accept ten percent of its own mass in accretion during an episode of RLOF. These models do not follow post-contact systems, so for those we assume 
that the primary is stripped down to its core mass, the secondary stays roughly the same mass as at the start of contact, and the period 
is governed by equation 3 as for the previous models. The pre-SN radius and post-SN lifetime of post-contact secondaries are then estimated 
from single stars of the corresponding mass, since in the majority of cases it has accreted only a small amount. 
Parameters of noninteracting systems are also taken from single star models (Dray \& Tout 2003). 

Under the assumptions used here, mergers are a frequent result of common envelope evolution. Unless the 
merger process itself involves significant asymmetric mass loss, the stars thus formed will not be runaway. 
Their subsequent evolution is likely also to differ from that of a normally-formed single star at their new 
mass. One might expect them to be rapidly-rotating and have non-ZAMS abundance profiles. However it is notable that 
many Blue Straggler stars have low rates of rotation, even though most scenarios for their creation involve mergers 
or significant amounts of accretion (Leonard \& Livio 1995, Sch{\"o}nberner \& Napiwotzki 1994). Probably 
merger products evolve more similarly to secondaries which have undergone accretion than to ZAMS stars. We calculate the O 
and WR lifetimes of these stars, therefore, by reference to post-accretion secondary models of the corresponding mass. It should be 
noted that these may not have exactly the same composition, so this comparison is relatively approximate; however, it 
is probably closer than assuming lifetimes appropriate to a ZAMS star of the new mass.

Systems for which the SN explosion prompts a merger of the new NS and its companion (mainly via the kick 
direction being such that the binary is significantly hardened to the point that the new periastron distance is less than 
the radius of the companion, rather than by direct collision) are fairly uncommon, happening to around 1 \% of systems which 
reach the SN stage. In this case a Thorne-{\. Z}ytkow object is formed (e.g. Podsiadlowski, Cannon \& Rees 1995). For systems 
which remain bound and close after the SN but do not merge immediately there is the possibility of unstable RLOF later to 
reach this same end. Whilst there 
has been some speculation that the unusual WN8 class of WR stars, which are runaways, may be Thorne-{\. Z}ytkow objects, 
we assume here that they will be short-lived and appear as red supergiants, thereby not affecting either the O or the WR statistics.

In panels {\bf c} and {\bf d} of Fig. 2 we consider a couple of other possible constraints on the kick distribution.
Whether or not supernovae which form BHs have strong kicks is a matter of debate (Jonker \& Nelemans 2004, Nelemans, Tauris  
\&  van den Heuvel 1999) but it is quite possible that their kick distribution is different 
from that of NS-forming explosions. Probably BHs formed by direct collapse do not have kicks, 
and it is possible that BHs formed by fallback have small kicks. 
A prototype for such a system may be Cygnus X-1 (Mirabel \& Rodrigues 2003), which 
contains a BH of around 10 solar masses with an 18 solar mass O supergiant 
companion, but seems not to have received any excess velocity from the supernova 
which formed the BH. This implies a SN with very little or no mass loss and no kick at 
a mass which is slightly lower than that generally assumed for direct collapse. Alternatively 
the kick may have been fortuitously directed so as to cancel out the velocity effect of mass loss.
In general, those studies which do indicate a difference in NS and BH kicks find lower or no kicks in BH-forming explosions.
In panel {\bf c} we assume that explosions which produce BHs (i.e. 
those of cores more massive than about $8 \, \msun$, Fryer 1999) have no kick at all.
Even without kicks, some mass is still lost in these supernovae and hence some of them still become runaways.
However the effect on the WR runaway population is fairly drastic, since WR secondaries are preferentially 
found with more massive primaries.

For panel {\bf d} we assume that systems with 
initial periods below ten days have small kicks (on the order of $30 \, {\rm km\,s}^{-1}$). 
This scenario is similar to that suggested to explain the apparently bimodal distribution of 
pulsar velocities (Pfahl et al. 2001, Podsiadlowski et al. 2004).
Systems with initially smaller periods will undergo RLOF earlier on in their evolution, revealing the core of the 
primary whilst it is still rotating rapidly, which may affect the kick magnitude.

The fractions of O and WR stars which will be observed as being runaway of course 
depends on the lifetimes those stars spend as runaways and the initial binary fraction.
Therefore also given in Fig. 2 are a number of properties of the resulting WR and O populations, including the 
observable runaway fractions ${\rm O_{r}}$ and ${\rm WR_{r}}$ (assuming an initially 100 per cent binary fraction), the 
fraction of systems which are unbound by the SN $f_{\rm unbound}$, the fraction of O or WR runaways which should have 
a NS or BH companion $f_{\rm r,binary}$ and the number ratio of WR stars to O stars. 
 Observationally in the Milky Way WR/O is between $0.1$ and $0.2$ (Maeder \& Meynet 1994). 
 As noted before, the parent population of BSS WR runaways is those binaries 
which are initially the most massive; even large SN kicks as used here frequently do not suffice to push them over 
the velocity limit to become runaway stars. Therefore the fraction of systems which experience a SN but 
remain bound may by very different to the fraction of runaways which remain bound.

\subsection{Comparison with observed values}

Whilst one may obtain similar runaway fractions for O and WR stars by restricting kicks as in panels {\bf c} and 
{\bf d} of Fig. 1, it is notable that such fractions are rather lower than the observed value. Without kick restrictions, 
similar O and WR runaway fractions are also obtained but they are somewhat larger. 
   In quantifying the observed runaway fractions, of course, it is important to be cautious about the 
completeness of runaway surveys. In the simplest case, that in which radial velocities alone are used to 
select for high velocity stars, as many as half of an isotropically-distributed sample of runaways will 
be missed (Cruz-Gonzalez et al. 1974). The study of Mason et al. (1998) accounts a star a runaway if 
it has absolute peculiar radial or space velocity greater than $30 {\rm kms}^{-1}$, or is further 
than 500 pc from the Galactic plane. That of Moffat et al. (1998) uses                               
a threshold of $42 {\rm kms}^{-1}$ in transverse velocity, which is equivalent to a threshold of  $30 {\rm kms}^{-1}$
in radial velocity alone. For WR stars in particular these values may also be affected by small number statistics. 
Nevertheless, since the Mason et al. study finds an O star runaway fraction of 8 percent and the Moffat et al. study 
finds a WR star runaway fraction of 9 percent, it is probably safe to say that the true fractions of stars with 
space velocity greater than $30 {\rm kms}^{-1}$ within these populations are similar, and that they are likely to be at 
the lower end of the 10 -- 20 percent range. 
  The runaway fractions we find without kick restrictions ({\bf b}) are then half or less of 
what is expected. These values are slightly higher than but broadly in agreement with those found
for O stars by Portegies Zwart (2000). If kicks for these systems are restricted, the runaway fractions of O and WR stars become similar, 
but are a factor of 5 -- 10 too low. 

There are a number of reasons why this might be the case. First, in real life, some runaways will arise 
from dynamical ejection. Following Hoogerwerf et al. (2001) we expect the number of these to be lower than 
those arising from the BSS, but they may be enough to make up some of the difference between the observed and 
theoretical values. As discussed above, it is difficult to predict whether the DES will produce differing 
runaway fractions for O and WR stars, but it is likely to be at least slightly more weighted towards making O star 
runaways than the BSS. Therefore one potential scenario is that there are no restrictions on SN kicks and the 
remainder of the runaways in both cases are made up by dynamical ejection. A further possibility is that the input 
parameters are incorrect. However additional simulations with a wide range of input distributions failed to 
produce a high enough runaway fraction for any reasonable set of parameters. We have also not considered triple and 
higher-order multiple systems. If nearly all O stars are born with companions, as seems likely, hierarchical 
triple systems may play an important -- and complex -- r{\^o}le in evolution and the production of runaway 
systems. In particular, there are three stars among the WR sample of runaways detected by Moffat et al. (1998) 
which must either be given their velocity by the DES or evolution in a multiple system, since they have O or B 
star companions. One of these is WR22, which, with a combined system mass of nearly eighty solar masses, probably requires 
dynamical interaction between a number of very massive stars to give it its velocity.

\begin{figure}
\vbox to140mm{\vfil
\psfig{figure=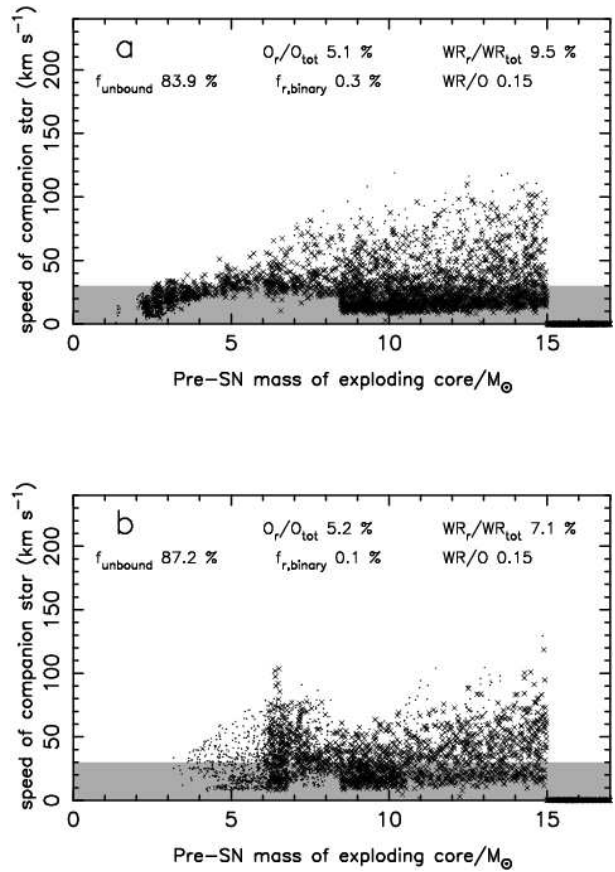,angle=0,width=80mm}
\caption{As Fig. 2, but with early lifetimes of O stars and low-mass WR-like stars excluded. Panel {\bf a} 
         shows a population with conservative mass transfer, and panel {\bf b} a population with non-conservative 
         mass transfer.}
\vfil}
\label{fig3}
\end{figure}

A pertinent question here is that of observability. Whilst we have calculated populations based on theoretical 
definitions of what constitutes an O star or a WR star, this does not take into account whether they can be 
detected as such or not. O stars are, on formation, hidden in dense molecular cloud cores (Heydari-Malayeri et al. 1999).
If the fraction of the early (non-runaway) O star lifetime during which they are not visible is large, this increases 
the observed O star runaway fraction significantly. Portegies Zwart (2000) estimates the average reduction in 
visible O star lifetime due to obscuration by their birth clouds as $10^{6}\,$years, which would only have a small effect.
Removing non-runaway O stars also increases the WR/O ratio, taking it further away from its observed value. 

  It is also true that not all of the stars which we have labelled as WR stars may be visible as such. Observationally, 
the WR phenomenon is strictly an atmospheric one, defined by low H and/or He abundances and broad emission 
lines. Stars which fit the theoretical definition of a WR star -- mainly, that its surface hydrogen abundance by 
mass be less than 0.4 -- may not display WR phenomena in their atmosphere if they are too cool or do not have a high enough 
mass-loss rate. Many of the quantities affecting the wind are related to the mass of the star, so it is convenient to take 
a mass limit below which a star with WR-like abundances will simply be observed as a helium star -- 
for example, studies such as that of Vanbeveren, Van Bever \& De Donder (1997) take this minimum mass to be $5\, \msun$. 
In the catalogue of van der Hucht (2001) the WR star listed as having the lowest determined mass is WR97, 
at $2.3\, \msun$. However the source paper for these values (Niemela, Cabanne \& Bassino 1995) quotes them 
as lower limits only, with the values corrected for inclination potentially being ten times greater.
The lowest confirmed WR mass is then $3.9\, \msun$ for WR31 (Lamontagne et al. 1996). The most evolved WC stars 
may also be shrouded by dust, appearing as faint red objects, although we do not consider that here since the 
lifetimes involved are short.
There are also a small number of stars in our models with WR-like abundances which become essentially cool He 
giants. These will be unable to maintain a WR atmosphere. We therefore rerun our simulations with the assumption that
stars with WR-like composition are not observable if they have masses below $ 3.9 \,\msun$ or ${\log \rm T_{eff}}$ 
below 4.0, and that massive stars younger than $10^{6}$ years are obscured. These are shown in Fig. 3; panel {\bf a} 
shows full evolutionary models with conservative mass transfer and these assumptions, and {\bf b} models with non-conservative mass 
transfer as above. 

These changes have a relatively small but positive effect. In the case that the runaway fraction amongst O stars 
is close to the Mason et al. (1998) value of 8 percent, the resulting O star runaway fraction is consistent with the 
conclusion of Hoogerwerf et al. (2000) that two-thirds of massive runaways are from the BSS and the remaining third from 
dynamical interaction. The WR star runaway fraction is similarly consistent with that of Moffat et al. (1998) -- in fact, an 
extra component from the DES is not needed at all in the case of conservative mass transfer.
Amongst the observed WR stars marked as runaways there is a higher incidence of non-compact companion stars than amongst 
runaway O stars, however, so it is likely that there is a significant DES component in the WR runaway population.

\subsection{Variation with metallicity}
A further variable which has the potential to have a strong effect, as noted above, is metallicity. Foellmi et al. (2003) find that eight
of their sample of 61 LMC WR stars have unusual radial velocities (i.e. significantly different from the 
mean value). This suggests a similar or slightly higher runaway fraction to the Milky Way. However, whereas several of 
the MW runaway sample have OB companions, all the suspected LMC runaways appear to be single. This could indicate a 
higher proportion of BSS runaways, but is most likely just a consequence of small number statistics. 
Lower-metallicity stars of similar mass and evolutionary status usually have smaller radii 
than comparable solar-metallicity stars. This leads to fewer 
contact systems, fewer mergers and more systems surviving until the first SN; and wind mass-loss rates are lower, 
which raises the mass limit above which a single star will go through 
a WR phase. WR secondaries are less affected by the lower mass-loss rates if their effective metallicity is raised by 
accretion. Lower mass loss in the wind of the primary also leads to a more massive primary at the 
time of explosion, which may result in a lower kick velocity. Therefore theoretically one would expect WR stars to 
have a higher runaway fraction at lower metallicity if the secondary accretes significantly (i.e. the conservative models) 
but otherwise to be relatively similar. Decreasing the metallicity also shifts the main sequence bluewards, increasing
non-runaway O star lifetimes. This is likely to reduce the fraction of the O star population which is runaway.  
These speculations are confirmed in Fig. 4, for which the same simulations as in Fig. 3 are carried out, but at 
Z = $0.01$. These general trends also held true when we ran the same simulations at Z = $0.004$.

\begin{figure}
\vbox to140mm{\vfil
\psfig{figure=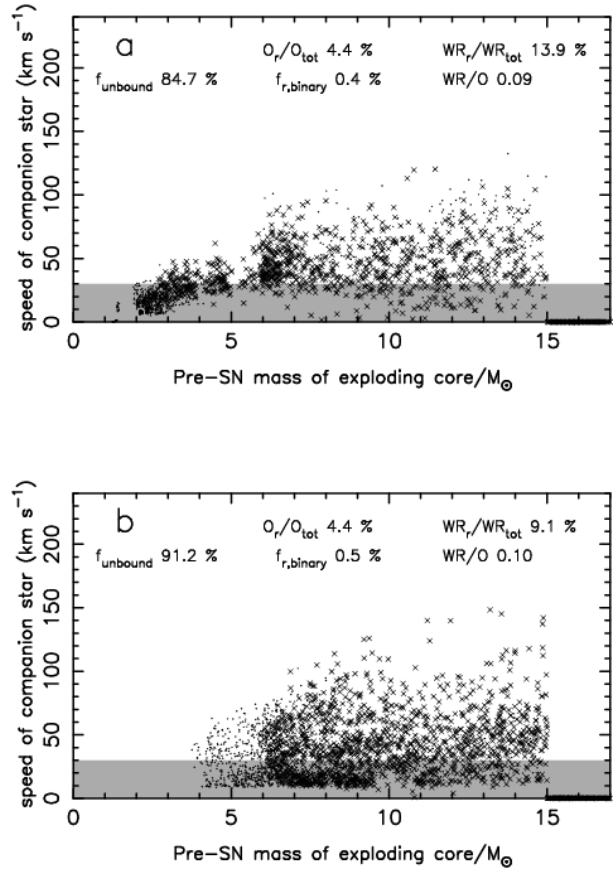,angle=0,width=80mm}
\caption{As Fig. 3, but for half solar metallicity.}
\vfil}
\label{fig4}
\end{figure}

From the above analysis it seems likely that the systems which have the potential to produce BSS WR runaways 
are those which have initially the most massive primary stars, and that because these primaries are still relatively 
massive when they explode, the velocities imparted to their companions are, on average, smaller, so that 
observed WR runaways represent the higher-velocity end of this spectrum.
This in turn increases the likelihood that such systems in fact remain bound and moving with 
relatively low velocity -- i.e. not enough to be registered as runaways. In the kick restriction 
scenarios above over fifty percent of secondaries which go on to become WR stars retain their 
companions, the vast majority of which do not attain runaway velocities.
Even without kick restrictions, the fraction of the total observed WR population which might be expected to have a 
compact companion, assuming no further interaction after the first SN, is around 8 percent. 

 The question which then arises is, where are the WR-BH 
and WR-NS post-explosion binaries which are predicted by this scenario? 

\section{Wolf--Rayet Binaries}

In the previous section we demonstrated that one consequence of a massive star 
population which is able to produce the expected numbers of runaways via 
supernovae in binary systems is that there is a non-negligible proportion 
of WR stars with compact companions. For the observed Galactic WR population 
of around 237, we would expect 
around 20 WR-compact object binaries, of which all but one or two will have BH 
companions. However, the observed Galactic WR + compact object population in 
fact has only one known member, Cygnus X-3. It is a matter of debate as to whether 
the compact object in this system is a BH or NS (e.g. Schmutz et al. 1996).
Cyg X-3 has a very tight orbit, with a period of only 4.8 hours. There is also a potential WR-BH 
binary in the metal-poor galaxy IC 10 (Bauer \& Brandt 2003), and one pulsar
with no optical counterpart, X1908+075, which displays signatures of orbiting in the wind of 
an unseen WR companion (Levine et al. 2004).

However, the typical WR--BH binary produced by our simulations is long-period (100 -- 1000 days) 
and has a relatively small eccentricity. Whilst there are some close systems, it seems 
unlikely that in the wide systems, which significantly outnumber them, there will be sufficient 
accretion to power X-ray emission. 

In addition, the dearth of obvious WR + BH binaries in the Galaxy is not surprising if we
examine how most stellar--mass black holes are found. By far the most
usual way (15 out of 18 known cases: McClintock \& Remillard, 2003) is
to measure the mass function of the companion star in a quiescent
transient and show that this exceeds the maximum neutron--star mass
$\simeq 3\msun$. If the system is not transient and thus has no
quiescent phase this is impossible, as the optically bright accreting
component prevents one obtaining a clean spectral line
radial--velocity curve for the companion. 

But WR + BH binaries are extremely unlikely to be transient. To power
an X--ray source (thus signalling the presence of a compact component)
by Roche lobe overflow or even by wind accretion requires a short
orbital period. The presence of a hot WR companion close to the
accretion disc guarantees that its temperature can never fall below
the hydrogen ionization value $\sim 6000$~K. This is for example why
none of the HMXBs is a conventional transient. 

The other method of finding black holes relies on rather indirect
arguments for HMXBs: one combines an absorption--line mass function
with a no--eclipse constraint, and gets a minimum value for the unseen
mass when all inclinations are allowed. There is in general no
guarantee that the resulting lower bound on the mass is tight enough
to require a black hole even if the required observations are
made. Thus it is unsurprising again that no second WR + BH binary has
been found.

The endpoints of the binary evolution are straightforward to
predict. If the secondary WR star leaves a neutron--star remnant in a tight
enough orbit (periods less than about 15~hr) this may coalesce via
gravitational radiation and possibly produce a gamma--ray burst as
well as a gravitational wave signal. If the binary was a runaway
the gamma--ray burst would presumably be rather distant from
star--forming regions, and so probably atypical. If the WR star leaves
a black hole, a similarly short orbital period would allow a BH + BH
coalescence, and a consequent gravitational wave source.

\section{O Star -- Compact Object Binaries}
One further consequence of evolutionary paths in which the binary is not separated by the SN is systems of O or B 
stars with compact companions. In particular, these systems may become visible as high mass X-ray binaries (HMXBs)
if material from the wind or from RLOF is accreted onto the compact component.

\begin{figure}
\vbox to140mm{\vfil
\psfig{figure=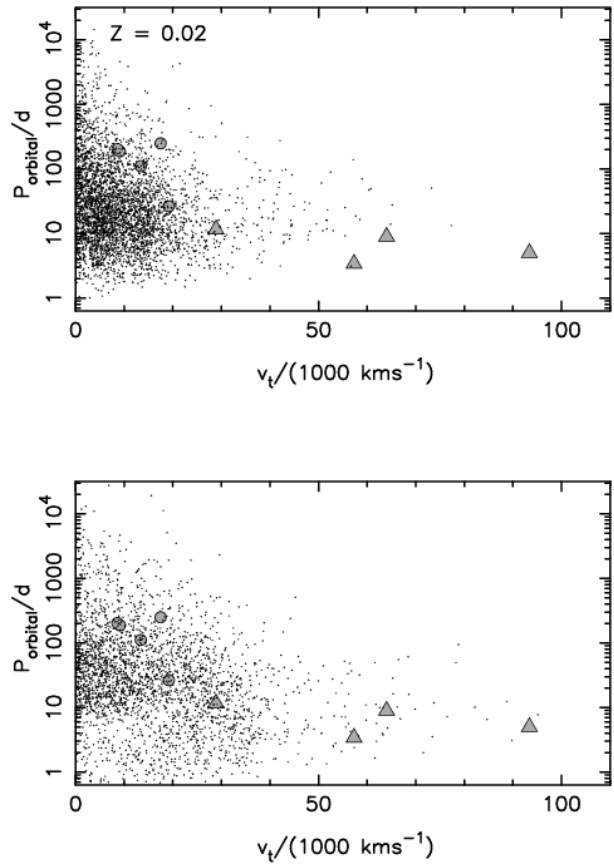,angle=0,width=80mm}
\caption{Transverse runaway velocities for systems which remain bound after the SN of the primary,
for conservative (upper panel) and non-conservative (lower panel) mass transfer. Large grey points
show the location of observed HMXB velocities from Chevalier \& Ilovaisky (1998) for systems with Be companions
(circles) and supergiant companions (triangles). Periods are obtained from the catalogue of Liu, van Paradijs
\& van den Heuvel (2000) and references therein.}
\vfil}
\label{fig5}
\end{figure}

Most HMXBs have small peculiar velocities. Chevalier \& Ilovaisky (1998) find for the Be X-ray binary systems an average 
transverse peculiar velocity $<v_{t}> = 11.3 \pm 6.7\,$km\,s$^{-1}$, i.e. not substantially different from that expected
for non-runaway O stars. Be X-ray binaries make up some 75 \% of the known HMXBs but, since they have transient emission, 
the real proportion is likely to be higher. The remaining HMXBs are persistently-emitting OB supergiant systems. These 
have larger velocities, probably related to a different evolutionary path:  $<v_{t}> = 42 \pm 14\,$km\,s$^{-1}$ (van 
den Heuvel et al. 2000). Other runaway O stars have indications of binarity and small mass functions which suggest they 
may have NS companions (e.g. Boyajian at al. 2005; McSwain et al. 2004). However in general searches for neutron star 
companions to runaway stars have suggested that the vast majority are single (Philp et al. 1996).  
   
From our simulations we find that around 5 -- 10 percent of O stars should have a compact companion of some sort, 
including direct collapse BHs. Many of these 
systems, like the WR systems mentioned above, will not support sufficient accretion to be X-ray bright. Of these 
binaries with compact companions, up to one percent have transverse peculiar velocities over $30 {\rm km\,s^{-1}}$.
This small fraction is unsurprising given that it is the systems with small kicks which are more likely to remain bound.
According to van Oijen (1989), there are around 960 -- 1850 O stars, 50 Be X-ray binaries and 
3 OB supergiant X-ray binaries within a distance of 2.5 kpc of the Sun. For this number of O stars, we would expect  
roughly 400 binaries consisting of a compact object plus any sort of companion, of which 4 or fewer will be runaway. 
Whilst the runaway compact object binaries we produce are close systems which might all be expected to be 
persistently X-ray bright, this number does seem rather small given that it is an upper limit.

In figure 5 we plot the observed period--transverse velocity distribution against that of systems 
which remain bound from our models. The smaller number ratio of observed to predicted low-velocity 
systems is consistent with these being transient Be X-ray binaries for which not all are visible at one 
time. The high-velocity X-ray binaries are notably easier to produce in the case of non-conservative 
mass transfer, for which there is a greater incidence of short-period binaries at the time of the first SN. 

We intend to explore this HMXB progenitor population further in a future paper.   

\section{Discussion}

As we have shown, the fraction of O stars and that of WR stars which are runaway may still 
be similar even if many of them come from supernova-separated binaries. This scenario is 
also consistent with the rarity of observed WR--NS binaries, provided that 
there is a population of WR--BH binaries which are not visible.

If we assume that, following the work of Hoogerwerf et al. (2001), both the DES and BSS 
operate in nature but that there is observational evidence that the BSS is responsible for 
about two-thirds of massive runaways, then there are a number of other conclusions which 
follow from this.

First, in order to be able to make enough runaways by this method, the vast majority of kick 
velocities must be large. This does not exclude 
X-ray binaries with low velocities since if kicks are randomly oriented in space then some will 
leave the system both bound and travelling with a relatively small velocity. Therefore the 
existence of some non-runaway HMXBs is unsurprising and expected in this scenario. 
However, if runaway O and WR stars are to be produced by the BSS in large enough numbers 
it is essential that at least some SNe in close binaries have large kicks, since these systems produce 
a significant fraction of the WR secondaries. Similarly our work supports the conclusions of 
Jonker \& Nelemans (2004) that BHs have kicks of similar magnitude to NSs, 
at least in the cases where they are formed by fallback. This is more consistent 
with a scenario where kicks are the result of anisotropic neutrino emission rather than 
anisotropic mass loss, since the amount of mass lost varies widely between BH-forming SN 
explosions.

This work also has some bearing on the mass limits above which 
stars collapse to BHs. If the core mass limit for direct collapse is much below 
$15 \, \msun$ then it is extremely hard to make enough runaway WR stars via the BSS.
Another scenario in which the number of BSS O and WR runaway stars becomes small is if 
the initial binary fraction amongst massive stars is significantly less than the 
here-assumed 100 percent, since initially-single stars 
are very unlikely to become runaway in either scenario. This suggests in turn that many observed 
single O and WR stars have been ejected from binaries either by the BSS or DES but have not attained 
enough velocity to count as runaways. From our models we note that the population of post-SN 
systems which do not have enough velocity to be accounted runaways has a larger binary fraction than 
the runaways but a smaller one than the parent population. This may be the origin of some of the 
observed field O stars which have not been designated runaways, but which have a binary fraction notably lower 
than for cluster O stars and higher than for O star runaways (Mason et al. 1998).

If the mass limit dividing cores which collapse to NSs and cores which collapse to 
BHs is much above $8 \, \msun$ then we would expect to see WR-NS binaries in our Galaxy. 
Since these are not observed, this argues that the mass limit cannot be any bigger than $8 \, \msun$. 
However, if it were much smaller, too many O--BH binaries and not enough O--NS binaries would be produced.   

A proper understanding of the ejection of stars from star clusters is also crucial to understanding the 
evolution of stellar systems. This applies \textit{a fortiori} to massive stars, since their feedback 
effects are thought to be responsible for unbinding and dispersing young bound star clusters 
(e.g. Hills 1980) and for regulating the rate and efficiency of star formation in such systems (e.g. McKee 1989).\\
\indent As noted previously, some $10\%-30\%$ of all O-type stars are runaways.  De Wit et al. (2004, 2005) and references 
therein estimate that an additional $20\%$ of O-type stars are not associated with any stellar cluster, but 
either have space velocities too small to be considered runaways, or have proper motions too small to 
allow their velocities to be reliably determined. This result, together with the fraction of O-stars which 
are defined as runaways implies that as many as half of all O-type stars have been expelled from their parent clusters.
For the calculations carried out here, up to 20 \% of O stars have been given a kick by binary interaction, 
although many of them have velocities too low to register officially as runaways. These, together with an extra 10 \% 
from the DES, could account for the homeless O stars.\\

\indent Feedback from O-type stars takes several forms, the most important of which on the typical lengthscales 
of star clusters ($\sim$pc) are photoionising radiation, stellar winds and supernovae. Ionising radiation and 
winds operate continuously for the duration of a star's lifetime, whereas a supernova is a single event occuring 
at the end of a star's main-sequence phase. The energy inputs from all three forms of feedback integrated over 
the stellar lifetime are approximately the same for massive stars ($\sim10^{51}$ erg for each feedback mechanism 
for a $30 \msun$ star - see Mac Low and Klessen (2004) for a discussion of the energy inputs from the three 
feedback mechanisms).\\
\indent If a star cluster is losing some of its complement of O-stars, the decrement in the total energy input 
to the cluster from stellar feedback is sensitive to the stage in the lifetime of each runaway O-star at which 
the star is expelled. Stars ejected by the DES are lost early in their lives and are thus unable to influence 
the evolution of their natal cluster by any means, whereas stars ejected by the BSS are usually at an advanced 
stage of their lives and will have already injected considerable quanitities of energy into the cluster by their 
winds and ionising radiation, although their supernovae will of course occur outside the cluster. 

If the results 
of this paper can be extended to \textit{all} homeless O-stars, $\sim10\%$ of massive stars are dynamically 
ejected from their parent clusters early in their main--sequence lifetimes, while $\sim20\%$ are ejected 
later in their lives by the supernova explosion of a binary partner. On the assumption that $10\%$ of O-stars 
are ejected near birth and a further $20\%$ just before exploding as supernovae and taking the time-integrated 
energy inputs from ionising radiation, winds and supernovae to be equal, a typical cluster may receive $17\%$ 
less energy from O-star feedback than if it retained all of its O-stars. This figure may often be an understimate 
since, as Hoogerwerf et al. (2001) point out, the interaction between AE Aurigae, $\mu$ Columbae and the binary 
$\iota$ Orionis removed $\sim70 \msun$ from the neighbourhood of the Trapezium cluster, comparable to the total 
mass of the Trapezium cluster itself. It is possible for dynamical interactions in a small-N system to expel 
N-2 objects and to leave only a tight binary (e.g. Kiseleva et al. 1998). A star cluster born with a rich 
population of O-stars could in principle be left with only two.\\
\indent Decreasing the energy input from feedback into a stellar cluster has the obvious consequence that the 
cluster is less likely to become unbound (if it was bound at formation). The efficiency of star formation will 
also be affected, although it is not clear whether it will be increased or decreased since feedback from 
massive stars can be both positive, in that it can induce or accelerate star formation locally, and negative, 
in that accretion onto existing stars can be halted and potentially-star-forming gas can be expelled from 
embedded stellar systems (e.g. Dale et al. 2005).\\

\section*{Acknowledgements} 

LMD and JED are supported by the Leicester PPARC rolling grant for
theoretical astrophysics, and RN by a PPARC Advanced Fellowship.  MEB
acknowledges the support of a UKAFF fellowship. ARK gratefully
acknowledges a Royal Society Wolfson Research Merit Award.

{\vspace{0.5cm}\small\noindent This paper
has been typeset from a \TeX / \LaTeX\ file prepared by the author.}

\label{lastpage}

\end{document}